\documentclass[assymb,11pt]{article}
\usepackage[pdftex,colorlinks=true,urlcolor=black,filecolor=black,linkcolor=black,
            pdftitle={String theory explanation of galactic rotation.},
            pdfauthor={Mark D. Roberts},
            pdfsubject={relativity,strings,galaxies},
            pdfkeywords={strings,scalar,tensor,galaxy,rotation},
            pagebackref,pdfpagemode=None,bookmarksopen=true]{hyperref}
\usepackage{amssymb}
\newcommand{\bc}{\begin{center}}
\newcommand{\ec}{\end{center}}
\newcommand{\bi}{\begin{itemize}}     
\newcommand{\ei}{\end{itemize}}
\newcommand{\bd}{\begin{description}} 
\newcommand{\ed}{\end{description}}
\newcommand{\bn}{\begin{enumerate}}   
\newcommand{\en}{\end{enumerate}}
\newcommand{\be}{\begin{equation}}
\newcommand{\ee}{\end{equation}}
\newcommand{\ber}{\begin{eqnarray}}
\newcommand{\ear}{\end{eqnarray}}
\newcommand{\ba}{\begin{array}}
\newcommand{\ea}{\end{array}}
\newcommand{\bv}{\begin{verbatim}}
\newcommand{\ev}{\end{verbatim}}

\newcommand{\Lg}{{\cal L}}

\begin{document}
\title{String theory explanation of galactic rotation.}
\author{
\href{http://www.violinist.com/directory/bio.cfm?member=robemark}
{Mark D. Roberts},\\
54 Grantley Avenue,  Wonersh Park,  GU5 0QN,  UK\\
mdr@ihes.fr
}
\date{$11^{th}$ of April 2010}
\maketitle
\bc
arXiv:
\href{http://arXiv.org/abs/1003.1309}
                      {\tt 1003.1309}
\ec
\begin{abstract}
The unique spherically symmetric metric which has vanishing weyl tensor,
is asymptotically desitter,  and can model constant galactic rotation curves is presented.
Two types of field equations are shown to have this metric as an exact solution.
The first is palatini varied scalar-tensor theory.
The second is the low energy limit of string theory modified by inclusion of
a contrived potential.
\end{abstract}
\section{Introduction}
Galactic rotation curves often exhibit speeds which are a constant independent of distance from
the center of the galaxy.
This is less than would be expected from solid body rotation where the rotation speed increases
with radial distance and more than would be expected from free orbit rotation where the speed
would decrease with radial distance.
Currently the majority view is that a large amount of 'dark matter'
occurs in non-luminous places to produce these rotation curves,
as opposed to the minority view which is that
constant rotation curves are caused by modified laws of gravity.
From a newtonian perspective the modification which works is the replacement of the reciprocal
potential by a logarthmic potential,
the spherically symmetric relativistic generalization of this \cite{mdrhs} has one free function
which in the present work is fixed by requiring that the weyl tensor vanishes.
This leaves the problem of finding which field equations the einstein tensor obeys and
both palatini varied scalar-tensor theory \cite{mdrfd}
and the low energy linit of string theory with added potential are found to work.
The usual method of approacing problems by starting with a lagrangian,
deriving field equations, finding solutions,
and then comparing with observations is thus reversed:
in the present case it is observation,  then metric,
then field equations and finally lagrangian.
\section{The metric}
Consider the line element
\be
\label{gal37}
ds^2=-(X+\lambda_t r^2)dt^2
+\frac{(X-v^2)^2}{X(X+\lambda_r r^2)}dr^2
+r^2d\Sigma_2^2,
\ee
where
\be
\label{Xang}
X\equiv(1+2v^2\ln(r)),~~~
d\Sigma_2^2\equiv d\theta^2+\sin(\theta)^2d\phi^2,
\ee
and $v$ is the constant speed of galactic rotation.
$g_{\theta\theta}$ and $g_{tt}$ are fixed upto a radial coordinate transformation by requiring
constant rotation curves,  $g_{rr}$ is fixed by requiring that the weyl tensor vanishes.
When $\lambda_t(\lambda_t-\lambda_r)=0$ the weyl tensor vanishes,
from now on only consider the solution $\lambda=\lambda_t=\lambda_r$.
Another property of the curvature is that the ricci scalar obeys a type of conformal
wave equation $R=6r\Box(1/r)$,
but there appears to be no pattern to the higher order ricci curvature invariants.
The curvature of the metric (\ref{gal37}) is characterized by the ricci scalar and
\be
\label{YZ}
Y\equiv R^t_{~t}-R^\theta_{~\theta}=\frac{2v^2}{r^2(X-v^2)},~~~
Z\equiv 3R^t_{~t}-R^{\theta}_{~\theta}=\frac{4v^6}{r^2(X-v^2)^3}.
\ee

The metric (\ref{gal37}) has vanishing weyl tensor so that it can be expressed in conformally
flat form $d\hat{s}^2=\Omega^2ds_{minkowski}^2$,  taking $\lambda=0$ and comparing the
$g_{\theta\theta}$ and $g_{rr}$ terms respectively
\be
\label{transformation}
\Omega^2R^2=r^2,~~~
R=\frac{Ar}{\sqrt{X}},
\ee
where $A$ is a constant of integration,
to invert the second of (\ref{transformation}) requires lambert W functions.
Together with (\ref{scalarfield}),  (\ref{transformation}) implies
\be
\label{explicitomega}
\Omega=\pm\frac{r}{R}=\pm\frac{1}{A}\sqrt{X}
=\pm\frac{1}{A}\exp\left(-\frac{\alpha\phi}{2}\right).
\ee
Thus the field equations (\ref{Sq}) can be expressed in terms of the conformal factor $\Omega$
and can be thought of as the difference in the ricci tensor between $d\hat{s}$ and $ds$.
This implies that any metric of the form $d\hat{s}^2=\Omega^2ds_{ricciflat}^2$ will also
be a solution of (\ref{Sq}).  In particular transforming the schwarzschild solution
\be
\label{gal79}
ds^2=
-X\left(1-\frac{2m\sqrt{X}}{r}\right)dt^2
+\frac{(X-v^2)^2}{X^2\left(1-\frac{2m\sqrt{X}}{r}\right)}dr^2
+r^2d\Sigma_2^2
\ee
and this is still a solution of (\ref{Sq}) with scalar field given by (\ref{scalarfield}).
\section{Palatini scalar-tensor theory}\label{pal}
Choosing the scalar field
\be
\label{scalarfield}
\phi=-\frac{1}{\alpha}\ln\left(X\right),
\ee
a tensor which vanishes is
\be
\label{Sq}
Q_{a b}\equiv R_{a b}-\alpha\phi_{;ab}+\frac{\alpha^2}{2}\phi_{a}\phi_{b}
-\frac{1}{2}g_{a b}\left\{\alpha\Box\phi+\alpha^2\left(\nabla\phi\right)^2-6\lambda\exp(\alpha\phi)\right\},
\ee
where $\alpha$ is an absorbable constant.
The field equations expressed in terms of the ricci tensor of palatini varied
scalar-tensor theory are
\be
\label{Spal}
8\pi\kappa^2 S_{ab}={\cal A}R_{ab}-{\cal A}'\phi_{;ab}-{\cal B_A}\phi_a\phi_b
-\frac{1}{2}g_{ab}\left\{{\cal A}'\Box\phi+{\cal A}''\left(\nabla\phi\right)^2+V(\phi)\right\},
\ee
where
\be
\label{BA}
{\cal B_A}={\cal B}+{\cal A}''-\frac{3}{2}\frac{{\cal A}'^2}{{\cal A}},
\ee
${\cal A}$ is called the primary dilaton function,  ${\cal B}$ is called the secondary
dilaton fuction and $V$ is called the potential.
The field equations (\ref{Sq}) are a particular case of the field equations (\ref{Spal}) with
\be
\label{particular}
{\cal A}=\exp(\alpha\phi),~~~
V=-6\lambda\exp(2\alpha\phi),~~~
{\cal B}=0,~~~
S_{ab}=0.
\ee
The field equations (\ref{Spal}) can be found by performing both metric and palatini variations
of the action
\be
\label{action}
S=\int d^4x\sqrt{-g}\left\{{\cal A}(\phi)R-{\cal B}(\phi)\left(\nabla\phi\right)^2-V(\phi)\right\}.
\ee
\section{Low energy string theory with a potential}\label{str}
The lagrangian for low energy string theory with a potential is
\be
\label{stlag}
\Lg=\exp\left(-2\phi\right)\left\{
R+\frac{\alpha}{2}\Box\phi-\beta\left(\nabla\phi\right)^2
-\frac{\gamma}{12}H^2-V\left(\phi\right)\right\},~~~
H^2\equiv H_{abc}H^{abc},
\ee
performing metric variation and then expressing the field equations in terms of the ricci tensor
\be
\label{stfld}
8\pi\kappa^2\exp\left(2\phi\right)S_{ab}=
R_{ab}-\alpha\phi_{;ab}+2\beta\phi_a\phi_b+\frac{\gamma}{2}H_{acd}H_{b..}^{~cd}
+g_{ab}\left(-\frac{\gamma}{6}H^2+V\right),
\ee
the metric (\ref{gal37}) is a solution with scalar field given by (\ref{scalarfield}) and
\be
\label{stringsol}
H^{abc}=\pm\sqrt{\frac{\alpha^2-4\beta}{2\gamma}}\epsilon^{abcd}\phi_d,~~~
V=\frac{v^2}{r^2(X-v^2)^3}\left(X(X-4v^2)+v^4\right).
\ee
\section{Properties and comments}\label{propertiessection}
Seventeen properties and comments follow.

Firstly seven points on the metric:\\
{\it M1} (\ref{gal37}) has constant rotation curves when $\lambda r$ is negligible,
the easiest way to see this is that the circular vector
\be
\label{prot}
P{^a}=\left(\frac{v}{r}\delta{^a}_{\theta}+\delta{^a}_{t}\right)f(r),
\ee
where $f$ is an arbitrary function of $r$ has acceleration
\be
\label{pprops}
\dot{P}_a=\lambda rf^2\delta^r_a,
\ee
so that when $\lambda$ vanishes the vector is acceleration free or geodesic.
For (\ref{gal79}) with the same vector (\ref{prot})
\be
\label{prop79}
\dot{P}_a=m\sqrt{X}(X-3v^2)\frac{f^2}{r^2}\delta^r_a,
\ee
so that when $m$ vanishes the vector (\ref{prot}) is again geodesic.
\\
{\it M2} (\ref{gal37}) is asymptotically desitter as $\lambda r^2$
increases much faster than $\ln(r)$.
For $\lambda=0$ the line element is not asymptotically flat as the $ln$ term in $g_{tt}$
diverges,  this is a problem with many models of galactic rotation which have no natural
long radial distance cut off,  having an asymptotically desitter spacetime provides such a cut off.
Whether this can be thought of as evidence of a non-vanishing cosmological constant or
just an indication of the effect of distant matter does not have to be chosen.\\
{\it M3} the short distance cut off for the metric is good,  for short distances (\ref{gal79})
shows that the galactic metric can approache schwarzschild spacetime.\\
{\it M4} at $r=1$,  $\ln(r)$ changes sign and it is not immediate where this is in meters,
however note that the solution is still a solution with $r\rightarrow r/r_o$ so that the
length scale $r_o$ is arbitrary and has to be fixed by other means.\\
{\it M5} it is not clear what,  if anything,  corresponds to the vanishing of the three metric
functions in (\ref{gal37}),  $(X-v^2)$ occurs in the denominator of curvature
invariants so as it approaches zero they diverge.\\
{\it M6} the line elements (\ref{gal37}) and (\ref{gal79})
were taken to be spherically symmetric rather than axi-symmetric,
but rotation of galactic spacetime would be expected,
spacetime rotation would need a more elaborate model,
one could simply choose $d\hat{s}^2=ds_{kerr}^2$ however kerr rotation is fundamentally short
range whereas galactic rotation is long range,
the simpler case of the newtonian model uses just the log potential $v^2\ln(r)$
so that the newtonian model is spherically symmetric
and this suggests that the simplest relativistic models are also spherically symmetric.\\
{\it M7} why choose a line element with vanishing weyl tensor in the first place:
from the perspective of the jordan formulation of einstein's equations one might expect
that at large distances the weyl scalar to be larger than the ricci scalar;
however from the perspective of the schwarzschild solution which has vanishing ricci tensor
and robertson-walker spacetime which has vanishing weyl tensor one might expect that vanishing
weyl tensor characterizes large distances,
then the question arises as to at what range the weyl tensor becomes non-negligible,
presumably this depends on the distance $r_o$ introduced above,
so far there the metric (\ref{gal79}) suggests that it is a universal length
rather than a length dependent on the mass of the galaxy under consideration.\\

Secondly four points on the palatini-scalar-tensor solution:\\
{\it P1} palatini variations work well for general relativity where they reproduce
the christoffel connection,  but not so for quadratic action theories where they act
back on the lagrangian to produces tensors which appear to be unconnected to anything:
for scalar-tensor theories they act back on the primary dilaton function producing a
non-christoffel conection and this turns out to be necessary in the present case;
purely metric variation of the action (\ref{action}) is unlikely to recover
the field equations (\ref{Sq}) as can be seen by subtracting off the $\theta,\theta$
component from the $t,t$ component:
for the palatini case the non-vanishing of $Y$ in (\ref{YZ})
can be matched to the $\phi_{;ab}$ term,  however for the purely metric case there is
no $\phi_{;ab}$ term.\\
{\it P2} the secondary dilaton function ${\cal B}$ is taken to vanish (\ref{particular}),
this means that there is no explicit kinetic term in (\ref{action}),
although there is an implicit kinetic term after palatini variation;
no explicit kinetic term is similar to some inflationary models
where the potential $V$ is related to the hubble constant and the kinetic term is less important.\\
{\it P3} as $k=0$ robertson-walker cosmology is conformally flat
the field equations (\ref{Sq}) are obeyed for any choice of the scale-factor,
however the field equations (\ref{Spal}) and  (\ref{particular})
might have as yet unexamined solutions and properties.\\
{\it P4} because the scalar-tensor theory involves palatini variations the underlying geometry
is no longer riemannian but rather weyl with object of non-metricity related to the primary
primary dilaton function ${\cal A}$,  see \cite{mdrfd}.

Thirdly three points on the properties of the string solution (\ref{stringsol}):\\
{\it S1} the values of the constants in (\ref{stringsol}) lead to a real three-form $H$,
they could have led to a complex one;
although there are three constants after absorbtion there is only one free parameter.\\
{\it S2} the potential in (\ref{stringsol}) is contrived,
perhaps with the inclusion of additional fields it will no longer be necessary.\\
{\it S3} the lagrangian (\ref{stlag}) includes a $\Box\phi$ term,
as for palatini-scalar-tensor theory this term is necessary in order for the field
equation involving $Y$ in (\ref{YZ}) to vanish,  however the term can be removed
from the lagrangian by integration by parts.

Fourthly three general comments:\\
{\it G1} the introduction of the absorbable constant $\alpha$ is done because in the better
known case of spherically symmetric static minimal scalar fields the value of the constant in
the equation analogous to (\ref{scalarfield}) depends on schwarzschild mass,
but the value of the constant in (\ref{Sq}) does not,
so it is anticipated that something similar could happen here.\\
{\it G2} the equations (\ref{Sq}) are assumed to be consistent as they can be lagrangian based,
however conservation equations and the initial value problem are not looked at:
conservation and euler equations are not immediate in palatini scalar-tensor theory,
also the initial value problems is not straightforward when there are
$\phi_{;ab}$ terms involved;
these consistency problems depend on the variables and connection
in which they are formulated.\\
{\it G3} it is not immediate how the result impacts other areas of physics,  for example
globular star clusters also exhibit unusual dynamics,
but usually have no overall $v$ as in the present case,
so far the post-newtonian approximation of (\ref{gal37}) has not been studied.\\
\section{Conclusion}
A relativistic model of galactic rotation curves was produced
which has the properties discussed in the last section.
A problem with the metric is how many field equations and lagrangians can it be a solution to.
The most important problem with the scalar-tensor theory is the physical origin of the scalar field
and the most likely explanation is that the scalar field comes from dimensional reduction.
The most important problem with the string theory solution is the contrived potential
which hopefully will at some time in the future be replaced by fields.
\section{Acknowledgement}
I would like to thank:
David Garfinkle for discussion on points P1 G1 G2 G3,
Tom Kibble for discussion on conformal invariance and point M5
and Arkady Tseytlin for discussion of point S3.

v1a 3march2010 | v1b 16march2010 | v1c 18march2010 | v1d 21march2010 | v1e 23march2010 |
v2a 11april2010
\end{document}